\documentstyle[11pt]{article}

\begin{document}

\input{epsf}

\def\beq{\begin{equation}}
\def\eeq{\end{equation}}
\def\bea{\begin{eqnarray}}
\def\eea{\end{eqnarray}}
\def\beas{\begin{eqnarray*}}
\def\eeas{\end{eqnarray*}}
\def\ov{\overline}
\def\ot{\otimes}

\newcommand{\hf}{\mbox{$\frac{1}{2}$}}
\def\sig{\sigma}
\def\De{\Delta}
\def\af{\alpha}
\def\be{\beta}
\def\la{\lambda}
\def\ga{\gamma}
\def\ep{\epsilon}
\def\vep{\varepsilon}
\def\half{\frac{1}{2}}
\def\third{\frac{1}{3}}
\def\fth{\frac{1}{4}}
\def\sth{\frac{1}{6}}
\def\tth{\frac{1}{24}}
\def\tde{\frac{3}{2}}

\def\zb{{\bar z}} 
\def\psib{{\bar \psi}} 
\def\etab{{\bar \eta }}
\def\gab{{\bar \ga}}
\def\vev#1{\langle #1 \rangle}
\def\inv#1{{1 \over #1}}

\def\CA{{\cal A}}       \def\CB{{\cal B}}       \def\CC{{\cal C}}
\def\CD{{\cal D}}       \def\CE{{\cal E}}       \def\CF{{\cal F}}
\def\CG{{\cal G}}       \def\CH{{\cal H}}       \def\CI{{\cal J}}
\def\CJ{{\cal J}}       \def\CK{{\cal K}}       \def\CL{{\cal L}}
\def\CM{{\cal M}}       \def\CN{{\cal N}}       \def\CO{{\cal O}}
\def\CP{{\cal P}}       \def\CQ{{\cal Q}}       \def\CR{{\cal R}}
\def\CS{{\cal S}}       \def\CT{{\cal T}}       \def\CU{{\cal U}}
\def\CV{{\cal V}}       \def\CW{{\cal W}}       \def\CX{{\cal X}}
\def\CY{{\cal Y}}       \def\CZ{{\cal Z}}

\newcommand{\np}{Nucl. Phys.}
\newcommand{\pl}{Phys. Lett.}
\newcommand{\prl}{Phys. Rev. Lett.}
\newcommand{\cmp}{Commun. Math. Phys.}
\newcommand{\jmp}{J. Math. Phys.}
\newcommand{\jpamg}{J. Phys. {\bf A}: Math. Gen.}
\newcommand{\lmp}{Lett. Math. Phys.}
\newcommand{\ptp}{Prog. Theor. Phys.}

\newif\ifbbB\bbBfalse                
\bbBtrue                             

\ifbbB   
 \message{If you do not have msbm (blackboard bold) fonts,}
 \message{change the option at the top of the text file.}
 \font\blackboard=msbm10 
 \font\blackboards=msbm7 \font\blackboardss=msbm5
 \newfam\black \textfont\black=\blackboard
 \scriptfont\black=\blackboards \scriptscriptfont\black=\blackboardss
 \def\Bbb#1{{\fam\black\relax#1}}
\else
 \def\Bbb{\bf}
\fi

\def\id{{1\! \! 1 }}
\def\bo{{\Bbb 1}}
\def\bI{{\Bbb I}}
\def\bC{{\Bbb C}} 
\def\bZ{{\Bbb Z}}
\def\CN{{\cal N}}

\title{New Integrable Models from Fusion}
\author{{\bf Z. Maassarani}\thanks{Work supported  by DOE grant no.
DE-FG02-97ER41027 and NSF grant no. DMR-9802813.} \\
\\
{\small Physics Department}\\
{\small University of Virginia}\\
{\small McCormick Rd,  Charlottesville, VA,  
22903 USA}\thanks{Email address: zm4v@virginia.edu} \\}
\date{}
\maketitle

\begin{abstract}
Integrable multistate or multiflavor/color models were 
recently introduced. They are generalizations of models corresponding to 
the   defining representations of the 
$\CU_q (\widehat{sl(m)})$ quantum algebras. 
Here I show that a similar generalization is possible for all higher 
dimensional representations.
The $R$-matrices and the Hamiltonians of these models are 
constructed by fusion. The $sl(2)$ case is treated in some detail and 
the spin-$0$ and spin-$1$ matrices are obtained in  explicit forms. 
This provides in particular a generalization of the 
Fateev-Zamolodchikov Hamiltonian. 
 
\end{abstract}
\vspace*{2.5cm}
\noindent
\hspace{1cm} February 1999\hfill\\

\thispagestyle{empty}

\newpage

\setcounter{page}{1}

\section{Introduction}

New one-dimensional integrable lattice models were introduced in \cite{amp},
within the framework  of the Quantum Inverse Scattering Method 
\cite{qism1,qism2,qism3}.
They correspond to a generalization whereby every state of the original 
model, for the defining representation of the Lie algebra $A_{m-1}$ 
(see for instance \cite{jimbo}), is replaced by an arbitrary number of copies. 
The structure of the $R$-matrices  is left unchanged by the
replacement, but a usual property, crossing unitarity, is lost. 
It is however still possible to construct integrable open boundary conditions.
The eigenvalue set of the transfer matrix is unchanged,  
while the degeneracies increase \cite{danzi}.  
For integrable periodic boundary conditions the 
eigenvalues and degeneracies change \cite{amp}. 

These models are of interest in some one-dimensional reaction-diffusion
processes \cite{adhr}. 
(Similar but different models were also considered in \cite{alcabar} 
as generalizations of the t-J model.) 
They also appear in connection   with the Hubbard model 
for electrons, as a specific limit \cite{aars,hakobian} or,
for $m=2$ and at a certain  value of the quantum parameter, as building
blocks for the natural generalizations of the Hubbard model \cite{ff,piezi}.
Upon fermionization, the multiple-states appear as different flavors of 
fermions. The general Hubbard models can then be seen as  multichannel
versions of the original model. 

All the `multiplicity' $A_{m-1}$ models of \cite{amp} correspond 
to the {\it defining} representations of the Lie algebras $A_{m-1}$.
As no quantum group formalism \cite{jimbo} 
is yet known for the multiplicity models,
it is not clear whether generalizations to higher 
dimensional representations are possible. 
There is no obvious or systematic way
to do  a  replacement of states by multiple copies in higher
dimensional $R$-matrices. The fusion method, known to work for ordinary
models, turns out to give the correct answer for the multistate models.  

The main result of this work is to show that such generalizations exist. 
I first review the fusion method for obtaining
higher dimensional solutions of the Yang-Baxter equation starting 
from an arbitrary solution. The results of section \ref{fusionsec}
are quite general and require only a minimal number of assumptions. 
This  method is then shown to work for the  models at hand. 
This yields  multiplicity models corresponding to some  higher 
dimensional representations, and also shows that iterated fusions
are possible. Thus  all  higher dimensional representations of $sl(m)$ 
have  multistate generalizations.   
Hamiltonians are obtained as the derivative of the new $\check{R}$-matrices.
I consider in detail  the spin-0 and spin-1 models  of $sl(2)$,
and give explicit expressions for their $R$-matrices.

\section{Fusion}\label{fusionsec}

The multiple fusion procedure for $sl(m)$  $R$-matrices was developed
in \cite{krs}. The idea for
other algebras is the same \cite{kr,kns}. A simple description was given 
in \cite{mene} for  two successive fusions. Here I review
the fusion method and give some additional general results.  

The models  are defined through their $R$-matrices. 
These are  solutions of the Yang-Baxter equation (YBE):
\beq
R_{12}(\la-\mu) \, R_{13}(\la) \, R_{23}(\mu) =
R_{23}(\mu) \, R_{13}(\la) \, R_{12}(\la-\mu) \label{ybe}
\eeq
Here and below, the notation $\CO_{ij}$ ($i \not= j$)
means that the operator $\CO$ acts non-trivially on the $i^{\rm th}$
and $j^{\rm th}$ spaces, and as the identity on the other spaces:
$\CO_{ij}=\sum_k \bI\otimes \cdots \otimes a_k^{(i)}\otimes \cdots \otimes
b_k^{(j)}\otimes \cdots \otimes \bI$ (if $i<j$),  
where $\CO =\sum_k a_k \otimes b_k$. 
Note also that the three spaces 1, 2 and 3 need {\it not} be copies
of the same space. 

Consider {\it any} solution $R$ of the Yang-Baxter equation (\ref{ybe}),
which becomes proportional to  a projector at some special value $\rho$ of 
the spectral parameter. Define the projector $\pi^{(1)}$ so 
that $\pi^{(1)}\propto R(\rho)$, and let
$\pi^{(2)}=\bI-\pi^{(1)}$
be the orthogonal complementary projector. 
Setting $\la-\mu = \rho$ in the YBE one obtains: 
\beq
\pi^{(1)}_{12}\,  R_{13}(\la)\, R_{23}(\la-\rho) =
R_{23}(\la-\rho)\, R_{13}(\la) \, \pi^{(1)}_{12} \label{ybep}
\eeq
Left and right multiplication of 
(\ref{ybep}) by $\pi^{(2)}_{12}$ yields two equations
\bea
& & \pi^{(1)}_{12} \, R_{13}(\la)\,  R_{23}(\la-\rho)\,  
\pi^{(2)}_{12} =0 \label{ybep1} \\
& & \pi^{(2)}_{12}\,  R_{23}(\la-\rho)\,  R_{13}(\la) \, 
\pi^{(1)}_{12}= 0 \label{ybep2}
\eea
Define two  fused matrices  by 
\bea
R^{(1)}_{<12>3}(\la) &=&
\pi^{(1)}_{12}\, R_{13}(\la)\, R_{23}(\la-\rho) \,\pi^{(1)}_{12}\label{fus1}\\
R^{(2)}_{<12>3}(\la) &=& 
\pi^{(2)}_{12}\, R_{13}(\la)\, R_{23}(\la-\rho) \,\pi^{(2)}_{12}\label{fus2}
\eea
Using equations (\ref{ybep1},\ref{ybep2}) and the YBE one  shows that the 
matrices (\ref{fus1}-\ref{fus2}) satisfy a YBE where one space is a tensor
product of two spaces:
\beq 
R^{(i)}_{<12>3}(\la-\mu) \, R^{(i)}_{<12>4}(\la) \, 
R_{34}(\mu) = R_{34}(\mu) \,
R^{(i)}_{<12>4}(\la)\,  R^{(i)}_{<12>3}(\la-\mu) 
\;\;,\;\;\; i=1,2 \label{ybefus1}
\eeq
Thus starting with a given solution of the YBE, we have obtained  
higher-dimensional $R$-matrices which are also solutions 
of the YBE. If $d_i, i=1,2,3$ are the dimensions of the spaces 1, 2 and 3,
then  $R^{(i)}_{<12>3}(\la)$ is a $d_1\times d_2 \times d_3$ dimensional 
square matrix. However 
the projection operators can be diagonalized simultaneously with
a change of basis matrix $S$. $S^{-1}_{12} R_{<12>3}^{(i)}(\la)\, 
S_{12}$ also satisfy  equation (\ref{ybefus1}), 
and their matrix expressions now contain
a  number of rows and columns with only vanishing elements, and  which can be 
deleted without spoiling the Yang-Baxter property. The remaining 
rows and columns can eventually be relabeled according to the 
states of the corresponding representations. 
Let ${\rm tr}( \pi^{(1)})= d$. Deleting the vanishing rows and columns from 
$S^{-1}_{12} R^{(1)}_{<12>3}(\la)\, S_{12}$ and 
$S^{-1}_{12} R^{(2)}_{<12>3}(\la)\, S_{12}$, yields 
a $d\times d_3$ and  $(d_1\times  d_2 -d)\times d_3$ dimensional 
square matrix, respectively. 

It is possible to fuse two matrices $R_{<12>3}^{(i)}(\la)$
to obtain a matrix $R_{<12><34>}^{(i)}(\la)$. Setting $\mu=\rho$ in 
(\ref{ybefus1}) and multiplying by the projection operators 
yields two equations similar to (\ref{ybep1},\ref{ybep2}). 
This leads to the following definitions:
\bea
R^{(1)}_{<12><34>}(\la) &=&
\pi^{(1)}_{34}\, R_{<12>4}^{(1)}(\la+\rho)\, R_{<12>3}^{(1)}(\la)
\,\pi^{(1)}_{34}\label{fus3}\\
R^{(2)}_{<12><34>}(\la) &=& 
\pi^{(2)}_{34}\, R_{<12>4}^{(2)}(\la+\rho)\, R_{<12>3}^{(2)}(\la) \,\pi^{(2)}_{34}\label{fus4}
\eea
Using the same methods as above one easily shows that these
matrices satisfy two Yang-Baxter equations ($i=1,2$):
\bea
&R^{(i)}_{<12><34>}(\la-\mu) \, R^{(i)}_{<12>5}(\la) \, 
R_{<34>5}^{(i)}(\mu) &\nonumber\\
& = R_{<34>5}^{(i)}(\mu)\, R^{(i)}_{<12>5}(\la)\,  
R^{(i)}_{<12><34>}(\la-\mu)&\label{ybefus2}\\
&R^{(i)}_{<12><34>}(\la-\mu) \, R^{(i)}_{<12><56>}(\la) \, 
R_{<34><56>}^{(i)}(\mu) &   \nonumber\\
&=  R_{<34><56>}^{(i)}(\mu)\, R^{(i)}_{<12><56>}(\la)\,  
R^{(i)}_{<12><34>}(\la-\mu) & \label{ybefus3}
\eea
Again $S_{34}^{-1} R^{(i)}_{<12><34>}(\la) \, S_{34}$, $i=1,2$, still
satisfy the above YBE's. Removing by hand 
a certain number of vanishing columns and rows, one
obtains matrices with smaller dimensions. 

Assume  now  that the original $R$-matrix is  regular and unitary, 
{\it i.e.}
\beq
R_{12}(0)= c\, \CP_{12}\;\;,\;\;\; R_{12}(\la)\, R_{21}(-\la)=
f(\la)\, \bI\, \;\;,\label{regun}
\eeq
where $\CP$ is the permutation operator. 
The function $f(\la)$ is then even, and $c$ is an arbitrary non-vanishing
complex number. 
The fused matrices (\ref{fus3},\ref{fus4}) inherit this property. 
It is however necessary to correctly normalize them  for the corresponding
value of $c$ not to vanish. This is  achieved  by the following  normalization: 
redefine $R_{<12><34>}^{(i)}(\la)$ as the right-hand side of 
(\ref{fus3},\ref{fus4}) multiplied by  $\left( f(\la+\rho) \right)^{-1}$. 
The reason for this normalization is the  vanishing of $f(\rho)$, for otherwise
the projector $R_{12}(\rho)$ would be invertible and therefore equal to 
the trivial identity projector $\bI$.  

The limit $\la\rightarrow 0$ for the 
redefined matrices (\ref{fus3}-\ref{fus4}) 
has to be taken with care.  Using the YBE one arrives at:
\bea
R^{(1)}_{<12><34>}(0) &=&  -c^2\, \CP_{13} \CP_{24}\, S_{12}^{-1}\pi^{(1)}_{12}
S_{12}\, S_{34}^{-1} \pi^{(1)}_{34} S_{34}\label{regs1}\\ 
R^{(2)}_{<12><34>}(0) &=&  +c^2\, \CP_{13} \CP_{24}\, S_{12}^{-1}\pi^{(2)}_{12}
S_{12}\, S_{34}^{-1} \pi^{(2)}_{34} S_{34}\label{regs2}\\
R^{(i)}_{<12><34>}(\la)  R^{(i)}_{<34><12>}(-\la) &=& \left[ f(\la)\right]^2
S_{12}^{-1} \pi^{(i)}_{12} S_{12}\,  
S_{34}^{-1} \pi^{(i)}_{34} S_{34}\;\;,\;\;\;
i=1,2\label{units}
\eea
where $R^{(i)}_{<34><12>}(\la)= \CP_{13} \CP_{24}\, R^{(i)}_{<12><34>}(\la)
\, \CP_{13}\CP_{24}$.

Having obtained  regular solutions of the YBE, the usual procedure for
constructing integrable spin-chain Hamiltonians with short-range interaction
is by taking logarithmic  derivatives of the transfer matrices at $\la=0$.
The quadratic hamiltonian density of such integrable hierarchies is 
nothing but the derivative at $\la=0$ of the matrix 
$\check{R}(\la)=\CP R(\la)$. Inclusion of  the normalizing factor gives:
\bea
\check{R}^{(i)}_{<12><34>}(\la)&=& \frac{1}{f(\la+\rho)} 
S_{12}^{-1} S_{34}^{-1} \pi_{12}^{(i)} \pi_{34}^{(i)} 
R_{32}(\la+\rho) \check{R}_{13}(\la)\\
& & \times\check{R}_{24}(\la) R_{23}(\la-\rho)
\pi_{12}^{(i)} \pi_{34}^{(i)} S_{12} S_{34}\;\; ,\; i=1,2 \nonumber
\eea
Taking the limit $\la\rightarrow 0$, I find:
\bea
\frac{d}{d\la}\check{R}^{(i)}_{<12><34>}(\la)_{|_0} &=&
- (\mp c^2)\frac{f''(0)}{2 f'(0)} \, S_{12}^{-1} \pi^{(i)}_{12}
S_{12}\, S_{34}^{-1} \pi^{(i)}_{34} S_{34} \nonumber\\
& &+ \frac{1}{2 f'(0)}  S_{12}^{-1} S_{34}^{-1} \pi^{(i)}_{12} \pi^{(i)}_{34}
\,\frac{d^2}{d\la^2}\left( R_{32}(\la+\rho) \check{R}_{13}(\la)
\right. \nonumber\\
& & \left.\times \check{R}_{24}(\la) R_{23}(\la-\rho)\right)_{|_0}\,
\pi^{(i)}_{12} \pi^{(i)}_{34} S_{12} S_{34} \label{hamden}
\eea 
The two signs are as in the regularity equations above.
The first term is proportional to the identity  and may be dropped.

It is in fact possible to fuse an arbitrary product of  $R$ matrices to obtain solutions of the YBE corresponding to most representations of a given 
Lie algebra. This was carried out in detail 
for $sl(2)$ and $sl(3)$  in \cite{krs,kr}, and also  works
for the  other algebras \cite{kns}. 

The above fusing scheme is now shown to work for the 
multiplicity $A_{m-1}$ models.  

\section{New integrable models} 

As seen above, apart from satisfying the Yang-Baxter equation,
the projector property is the only additional ingredient
needed to construct fused matrices. In particular $R$ does not 
have to correspond to a smallest representation to be able to 
apply the fusion method. 
The degeneration of the generically invertible  $R$-matrix  
to a projector, for a certain value of the spectral parameter, is not 
automatic. It is however a quite common property, especially for
the $R$-matrices based on Lie algebras. 
For the models studied here, a quantum group structure does not
seem to exist, and one has to verify that a projector point exists. 

I now apply this formalism to the models of \cite{amp}. This 
will yield new matrices which define new integrable hierarchies. 
The models of \cite{amp} are defined  as follows.
Take positive integers  $n_i$ and  $m$ such that 
\beq
\sum_{i=1}^m n_i =n \;\;\;\;\;\;{\rm and}
\;\;\;\; 1\leq n_1 \leq ... \leq n_m \leq n -1
\eeq
The inequality restrictions avoid multiple counting of models, but
can otherwise be relaxed.
The set of  $n$ basis states  is the disjoint union of $m$  sets $\CA_i$:
\beq
{\rm card}\,(\CA_i) = n_i \;\;,\;\;\; \CA_i \cap  \CA_j= \emptyset 
\;\;\;\;{\rm for}\;\;\; i\not= j 
\eeq
$\CA_i$ should not be confused with the Lie algebra $sl(i+1)$. 
Let $E^{\af\be}$ be a square matrix  with a one at row $\af$ 
and column $\be$ and zeros otherwise.
Define   the following operators:
\bea
\tilde{P}^{(+)}&=& \sum_{1\leq i< j \leq m}\sum_{\af_i \in \CA_i}
\sum_{\af_j\in \CA_j} E^{\af_j\af_i}\otimes E^{\af_i\af_j} \label{pm}\\
\tilde{P}^{(-)}&=& \sum_{1\leq i< j \leq m}\sum_{\af_i \in \CA_i}
\sum_{\af_j\in \CA_j} E^{\af_i\af_j}\otimes E^{\af_j\af_i} \label{pp}\\
\tilde{P}^{(3)}&=&\sum_{1\leq i< j \leq m}\;\sum_{\af_i \in \CA_i}
\sum_{\af_j\in \CA_j} \left(x E^{\af_j\af_j}\otimes E^{\af_i\af_i}
+ x^{-1} E^{\af_i\af_i}\otimes E^{\af_j\af_j}\right) \\
\eea
The twist parameters were taken equal to one arbitrary complex parameter $x$.
Let $y=e^{i\la}$ and $q=e^{i\gamma}$, where $\la$ is the spectral 
parameter and $\ga$ the quantum  parameter.
The $R$-matrix  is then given by:
\bea
&R(\la)= \CP\, \sin(\la+\ga) + \tilde{P}\, \sin\la &\label{ram}\\
&\tilde{P}\equiv \tilde{P}^{(3)} -(q^{-1}\, \tilde{P}^{(+)} + 
q\, \tilde{P}^{(-)}) &
\eea
This model is  denoted by $(n_1, ... , n_m;m,n)$.
For $n=m$  and $x=1$
one obtains  the $A_{m-1}$ $R$-matrix of \cite{jimbo}.
$R(\la)$ is  regular and unitary:
\bea
&R_{12}(0)=\CP_{12}\,\sin\ga &\label{regu}\\
& R_{12}(\la)\, R_{21}(-\la)=\bI\, 
f(\la)=\bI\,\sin(\ga+\la) \sin(\ga-\la) &\label{unit}
\eea
where $R_{21}(\la)=\CP_{12}\, R_{12}(\la)\, \CP_{12}$. As seen in section
\ref{fusionsec}, these properties are 
inherited by some of the  fused matrices. 

The right-hand side of   (\ref{unit}) shows that $\la=\pm \ga$
are possible projector points. Further checks show that
matrix (\ref{ram}) yields a projector at $\la=\rho=-\gamma$:
$R(-\gamma)= -\tilde{P}\sin\gamma $. Let  
\beq
\pi^{(1)}=\frac{1}{x+x^{-1}} \tilde{P} \;\; ,\;\;\; 
\pi^{(2)}=\bI-\pi^{(1)} \label{proj}
\eeq
One has: $(\pi^{(i)})^2=\pi^{(i)}$, $i=1,2$,  and 
$\pi^{(1)}\pi^{(2)}=\pi^{(2)}\pi^{(1)}=0$.
The dimensions of these projectors are given by their traces:
\bea
{\rm tr}\,(\pi^{(1)}) &=& \sum_{i<j} n_i \, n_j = \frac{1}{2} \left( 
n^2- \sum_i (n_i)^2 \right) \\
{\rm tr}\,(\pi^{(2)}) &=& n^2 - \sum_{i<j} n_i\,  n_j = \frac{1}{2} \left( 
n^2+ \sum_i (n_i)^2 \right)
\eea
The matrix $S$ which diagonalizes $\pi^{(1)}$ and $\pi^{(2)}$ is given by:
\bea
S&=& \sum_{i=1}^{m} \sum_{\af_i \in \CA_i} \sum_{\beta_i \in \CA_i}
E^{\af_i\af_i}\otimes E^{\beta_i\beta_i}\nonumber\\
& & +\sum_{1\leq i< j \leq m}\;\sum_{\af_i \in \CA_i}
\sum_{\af_j\in \CA_j} \left( E^{\af_i\af_i}\otimes E^{\af_j\af_j}
-\frac{x}{q} E^{\af_j\af_j}\otimes E^{\af_i\af_i}\right)\nonumber\\
& &+\sum_{1\leq i< j \leq m}\sum_{\af_i \in \CA_i}
\sum_{\af_j\in \CA_j}\left( E^{\af_i\af_j}\otimes E^{\af_j\af_i}
+\frac{1}{x q} E^{\af_j\af_i}\otimes E^{\af_i\af_j}\right)\label{ss}\\
S^{-1}&=&\sum_{i=1}^{m} \sum_{\af_i \in \CA_i} \sum_{\beta_i \in \CA_i}
E^{\af_i\af_i}\otimes E^{\beta_i\beta_i}\nonumber\\
& & +\frac{q}{x+x^{-1}}\sum_{1\leq i< j \leq m}\;\sum_{\af_i \in \CA_i}
\sum_{\af_j\in \CA_j} \left( \frac{x}{q}E^{\af_i\af_i}\otimes E^{\af_j\af_j}
- E^{\af_j\af_j}\otimes E^{\af_i\af_i}\right)\nonumber\\
& & +\frac{q}{x+x^{-1}}\sum_{1\leq i< j \leq m}\sum_{\af_i \in \CA_i}
\sum_{\af_j\in \CA_j}\left( E^{\af_i\af_j}\otimes E^{\af_j\af_i}
+\frac{1}{x q} E^{\af_j\af_i}\otimes E^{\af_i\af_j}\right)\label{ss1}
\eea
The diagonalized projectors then read
\bea
S^{-1}\, \pi^{(1)}\, S &=& \sum_{1\leq i< j \leq m}\;\sum_{\af_i \in \CA_i}
\sum_{\af_j\in \CA_j}  E^{\af_j\af_j}\otimes E^{\af_i\af_i}\\
S^{-1}\, \pi^{(2)}\, S &=&  \sum_{i=1}^{m} \sum_{\af_i \in \CA_i} 
\sum_{\beta_i \in \CA_i} E^{\af_i\af_i}\otimes E^{\beta_i\beta_i}\nonumber\\
& & +\sum_{1\leq i< j \leq m}\;\sum_{\af_i \in \CA_i}
\sum_{\af_j\in \CA_j}  E^{\af_i\af_i}\otimes E^{\af_j\af_j}
\eea
The explicit expressions of the four fused matrices $R^{(i)}_{<12>3}$ and
$R^{(i)}_{<12><34>}$ can be found by straightforward if tedious 
expansions of the products in (\ref{fus1},\ref{fus2},\ref{fus3},\ref{fus4}),
using the explicit expressions (\ref{ram},\ref{proj},\ref{ss},\ref{ss1}).  
Similarly, the quadratic hamiltonian density 
is obtained by replacing the matrices in (\ref{hamden}) and expanding
the products. 

The following point is worth mentioning. 
Since $R_{12}(-\ga) \, R_{21}(\ga) =0$, it may seem that 
the complementary projector $\pi^{(2)}$ is proportional to $R_{21}(\ga)$.
However this is generically {\it not} the case. 
It is true only when all the $n_i$ are equal to one and $x=q^{\pm 1}$. 
This is another distinguishing feature of the $n_i \not= 1$ models.

\section{Some $sl(2)$ models}\label{sl2}

Consider now the   $m=2$ case, {\it i.e.} the XXC models \cite{xxc}. 
They have an underlying $sl(2)$ structure.  Their $R$ matrix 
is just the multistate version of the one  corresponding to 
the spin-$\frac{1}{2}$ model $(n_1=n_2=1)$. 

Tedious but simple calculations lead to the 
following matrix which carries  spin-0$\times$spin-$\frac{1}{2}$:
\bea
R_{<12>3}^{(1)}(\la)&=& \sin\la \;\sin
(\la+2\gamma)\times \label{rm012}\\
& &\left(\sum_{\af_1,\be_1} \sum_{\af_2}
x\, E^{\af_2\af_2} \otimes E^{\be_1\af_1}\otimes E^{\af_1\be_1}
+\sum_{\af_1}\sum_{\af_2,\be_2} x^{-1} E^{\be_2\af_2}\otimes E^{\af_1\af_1}
\otimes E^{\af_2\be_2} \right)\nonumber
\eea
The dimension of this matrix is $n_1\, n_2 \, n$, as expected. 
The corresponding two-dimensional vertex model has $n_1\, n_2$ possible
states on, {\it e.g.}, the horizontal 
links and $n$ states on the vertical links.
 
Note that the $x+x^{-1}$ denominator has dropped out of the final result and
therefore $R$ is defined for all finite values of $x$, including $x=\pm i$.
This also holds for the three  fused matrices given below.

The matrix which carries  spin-0$\times$spin-0 can then be obtained:
\bea
R_{<12><34>}^{(1)}(\la)&=& \frac{\sin(\la-\gamma) 
\sin(\la+\gamma)\sin(\la+2\gamma)}{\sin(2\gamma-\la)}
\sum_{\af_1,\be_1} \sum_{\af_2,\be_2} 
E^{\af_2\be_2}\otimes  E^{\af_1\be_1}\otimes 
E^{\be_2\af_2}E^{\be_1\af_1}\nonumber\\
& =& \frac{\sin(\la-\gamma) 
\sin(\la+\gamma)\sin(\la+2\gamma)}{\sin(2\gamma-\la)}\;
\CP_{13}^{(\CA_2)} \,\CP_{24}^{(\CA_1)}\label{rm00}
\eea
$\CP_{13}^{(\CA_2)}$ and  $\CP_{24}^{(\CA_1)}$ are the permutation
operators in the subspaces $\CA_2$ and $\CA_1$, respectively. 
The dimension of this matrix is $(n_1 n_2)^2$, as it should be.
It satisfies the regularity and unitarity properties (\ref{regs1},\ref{units}).
Here the vertex model has $n_1\, n_2$ possible states on both
the horizontal and vertical links.
The resulting spin-chain is however rather trivial as the $\la$-dependence can
be normalized away, and the operator part yields $S_{12}^{-1}\pi_{12}^{(1)}
S_{12}\, S_{34}^{-1}\pi_{34}^{(1)} S_{34}$, {\it i.e.} 
the identity operators in  the spin-0 subspaces. 
 
The matrix carrying spin-1$\times$spin-$\frac{1}{2}$ is:
\bea
\frac{1}{\sin(\la+\gamma)}\, R_{<12>3}^{(2)}(\la) & =& +\sin(\la+2\gamma)
\sum_{\af_1,\be_1,\ga_1} E^{\af_1\be_1}\otimes 
E^{\ga_1\af_1}\otimes E^{\be_1\ga_1}\label{rm112}\\
& & + \sin(\la+2\gamma) \sum_{\af_2,\be_2,\ga_2}  
E^{\af_2\be_2}\otimes E^{\ga_2\af_2}\otimes E^{\be_2\ga_2}\nonumber\\
& & +y \,\sin(2\gamma) \sum_{\af_1}\sum_{\af_2,\be_2}  
E^{\af_2\af_1}\otimes E^{\be_2\af_2} \otimes E^{\af_1\be_2}\nonumber\\
& & +q\, y \,\sin\gamma \, \sum_{\af_1,\be_1}\sum_{\af_2}  
E^{\be_1\af_1}\otimes E^{\af_2\be_1} \otimes E^{\af_1\af_2}\nonumber\\
& & + x\, \sin(\la+\gamma) \sum_{\af_1,\be_1}\sum_{\af_2}  
E^{\be_1\af_1}\otimes E^{\af_2\af_2}\otimes  E^{\af_1\be_1}\nonumber\\
& & + x^{-1} \sin(\la+\gamma) \sum_{\af_1}\sum_{\af_2,\be_2}  
E^{\af_1\af_1}\otimes E^{\be_2\af_2}\otimes E^{\af_2\be_2}\nonumber\\
& & +x \, y^{-1} \sin \gamma \, \sum_{\af_1}\sum_{\af_2,\be_2}  
E^{\af_1\af_2}\otimes E^{\be_2\be_2} \otimes E^{\af_2\af_1}\nonumber\\
& & + x^{-1} q^{-1} y^{-1} \sin(2\gamma) \sum_{\af_1,\be_1}\sum_{\af_2}  
E^{\af_1\af_1}\otimes E^{\be_1\af_2}\otimes  E^{\af_2\be_1}\nonumber\\
& & + x^2 \sin\la \, \sum_{\af_1}\sum_{\af_2,\be_2}  
E^{\af_2\af_2}\otimes E^{\be_2\be_2} \otimes E^{\af_1\af_1}\nonumber\\
& & + x^{-2} \sin\la\, \sum_{\af_1,\be_1}\sum_{\af_2}  
E^{\af_1\af_1}\otimes E^{\be_1\be_1}\otimes E^{\af_2\af_2}\nonumber
\eea
The vertex model model has $(n_1)^2 + (n_2)^2 +n_1\, n_2$ states on 
the horizontal links and $n$ states on the vertical links. 

The preceding matrix is then used to find the spin-1$\times$spin-1 matrix:
\bea
\frac{\sin(2\gamma-\la)}{\sin(\la+\gamma)}&\times & 
R_{<12><34>}^{(2)}(\la) =\label{rm11}\\
& & +\sin(\la+\gamma) \sin(\la+2\gamma)
\sum_{\af_1,\be_1,\ga_1,\delta_1} E^{\af_1\be_1}\otimes 
E^{\ga_1\delta_1}\otimes E^{\be_1\af_1}\otimes 
E^{\delta_1\gamma_1}\nonumber\\
& & +\sin(\la+\gamma) \sin(\la+2\gamma)
\sum_{\af_2,\be_2,\ga_2,\delta_2} E^{\af_2\be_2}\otimes 
E^{\ga_2\delta_2}\otimes E^{\be_2\af_2}\otimes E^{\delta_2\gamma_2}\nonumber\\
& & +y \sin(2\gamma)\sin(\la+\ga)
\sum_{\af_1}\sum_{\af_2,\be_2,\ga_2} E^{\af_2\af_1}\otimes 
E^{\be_2\gamma_2}\otimes E^{\af_1\af_2}\otimes E^{\gamma_2\be_2}\nonumber\\
& & +y^{-1} \sin(2\gamma)\sin(\la+\ga)
\sum_{\af_1,\be_1,\gamma_1}\sum_{\af_2} E^{\af_1\be_1}\otimes 
E^{\gamma_1\af_2}\otimes E^{\be_1\af_1}\otimes E^{\af_2\gamma_1}\nonumber\\
& & +y \sin(2\gamma)\sin(\la+\ga)
\sum_{\af_1,\be_1,\gamma_1}\sum_{\af_2} E^{\af_1\be_1}\otimes 
E^{\af_2\gamma_1}\otimes E^{\be_1\af_1}\otimes E^{\gamma_1\af_2}\nonumber\\
& & +y^{-1} \sin(2\gamma)\sin(\la+\ga)
\sum_{\af_1}\sum_{\af_2,\be_2,\ga_2} E^{\af_1\af_2}\otimes 
E^{\be_2\gamma_2}\otimes E^{\af_2\af_1}\otimes E^{\gamma_2\be_2}\nonumber\\
& & +(\sin\gamma \sin(2\gamma)+\sin\la\sin(\la+\ga))
\sum_{\af_1,\be_1}\sum_{\af_2,\be_2} E^{\af_1\be_1}\otimes 
E^{\af_2\be_2}\otimes E^{\be_1\af_1}\otimes E^{\be_2\af_2}\nonumber\\
& & +y^2 \sin\gamma\sin(2\gamma)
\sum_{\af_1,\be_1}\sum_{\af_2,\be_2} E^{\af_2\af_1}\otimes 
E^{\be_2\be_1}\otimes E^{\af_1\af_2}\otimes E^{\be_1\be_2}\nonumber\\
& & +y^{-2} \sin\gamma\sin(2\gamma)
\sum_{\af_1,\be_1}\sum_{\af_2,\be_2} E^{\af_1\af_2}\otimes 
E^{\be_1\be_2}\otimes E^{\af_2\af_1}\otimes E^{\be_2\be_1}\nonumber\\
& & +2 x \, q^{-1} y\cos\gamma \sin(2\gamma)\sin\la\,
\sum_{\af_1,\be_1}\sum_{\af_2,\be_2} E^{\af_2\af_1}\otimes 
E^{\be_2\af_2}\otimes E^{\af_1\be_1}\otimes E^{\be_1\be_2}\nonumber\\
& & +x^{-1} q\, y \sin\gamma\sin\la\,
\sum_{\af_1,\be_1}\sum_{\af_2,\be_2} E^{\af_1\be_1}\otimes 
E^{\af_2\af_1}\otimes E^{\be_1\be_2}\otimes E^{\be_2\af_2}\nonumber\\
& & +x^2 \sin\la\sin(\la+\gamma)
\sum_{\af_1,\be_1,\gamma_1}\sum_{\af_2} E^{\af_1\be_1}\otimes 
E^{\af_2\af_2}\otimes E^{\be_1\gamma_1}\otimes E^{\gamma_1\af_1}\nonumber\\
& & +x^2 \sin\la\sin(\la+\gamma)
\sum_{\af_1}\sum_{\af_2,\be_2,\gamma_2} E^{\af_2\be_2}\otimes 
E^{\gamma_2\af_2}\otimes E^{\af_1\af_1}\otimes E^{\be_2\gamma_2}\nonumber\\
& & +x^{-2} \sin\la\sin(\la+\gamma)
\sum_{\af_1,\be_1,\gamma_1}\sum_{\af_2} E^{\af_1\be_1}\otimes 
E^{\gamma_1\af_1}\otimes E^{\be_1\gamma_1}\otimes E^{\af_2\af_2}\nonumber\\
& & +x^{-2} \sin\la\sin(\la+\gamma)
\sum_{\af_1}\sum_{\af_2,\be_2,\gamma_2} E^{\af_1\af_1}\otimes 
E^{\af_2\be_2}\otimes E^{\be_2\gamma_2}\otimes E^{\gamma_2\af_2}\nonumber\\
& & +x^3 q \, y^{-1} \sin\gamma\sin\la\,
\sum_{\af_1,\be_1}\sum_{\af_2,\be_2} E^{\af_1\af_2}\otimes 
E^{\be_2\be_2}\otimes E^{\be_1\be_1}\otimes E^{\af_2\af_1}\nonumber\\
& & +2 x^{-3} q^{-1} y^{-1} \cos\gamma \sin(2\gamma)\sin\la\,
\sum_{\af_1,\be_1}\sum_{\af_2,\be_2} E^{\af_1\af_1}\otimes 
E^{\be_1\af_2}\otimes E^{\af_2\be_1}\otimes E^{\be_2\be_2}\nonumber\\
& & +x^4 \sin(\la-\gamma)\sin\la\,
\sum_{\af_1,\be_1}\sum_{\af_2,\be_2} E^{\af_2\af_2}\otimes 
E^{\be_2\be_2}\otimes E^{\af_1\af_1}\otimes E^{\be_1\be_1}\nonumber\\
& & +x^{-4} \sin(\la-\gamma)\sin\la\,
\sum_{\af_1,\be_1}\sum_{\af_2,\be_2} E^{\af_1\af_1}\otimes 
E^{\be_1\be_1}\otimes E^{\af_2\af_2}\otimes E^{\be_2\be_2}\nonumber
\eea
This matrix does satisfy  the regularity and unitarity properties  
(\ref{regs2},\ref{units}).
The number of states on both the horizontal and vertical links is
now $(n_1)^2 + (n_2)^2 +n_1\, n_2$. 

The four matrices (\ref{rm012},\ref{rm00},\ref{rm112},\ref{rm11}) have
2, 1, 10, 19 types of terms, respectively. This was expected from the 
$S^z$ conservation  of $sl(2)$. 
When $n_1=n_2=1$  the matrices give the
$sl(2)$ 2-, 1-, 10-, 19-vertex models. The latter two models can be found 
in \cite{fzmod,ppny,mene2}. To carry out a comparison 
it is necessary to relabel
the states so that $R$ looks like an operator acting on the tensor
product of two spaces: $R=\sum E\otimes E$. 
For instance, in (\ref{rm112}), with $\CA_1=\{ 1\}$, $\CA_2=\{ 2\}$, one can
take the matrix element  $\sin(\la+\gamma) \sin(\la+2\gamma) E^{11}\otimes
E^{11}\otimes E^{11}$ to correspond to $|+1\rangle |+\frac{1}{2}\rangle\times
|+1\rangle |+\frac{1}{2}\rangle$, and 
$(x\, q\, y)^{-1}\sin(2\gamma)\sin(\la+\gamma) E^{11}\otimes E^{12}\otimes
E^{21}$ to $|+1\rangle |-\frac{1}{2}\rangle\times
|0\rangle |+\frac{1}{2}\rangle$. 
The matrix (19) of reference \cite{mene2}
can be obtained by fusing the symmetric form of the $sl(2)$ $R$-matrix
\cite{xxc}. It is also necessary to do a gauge transformation after fusing,
in order to render the resulting matrix completely symmetric. 
This introduces the unusual square-root element $d$ of that
reference. Finally, let  $u=\la+\eta/2$ and $\eta=\gamma$ to complete the 
identification. Similar remarks apply to the identification
of the 19-vertex models. 

The interpretation of the multiple-states in terms of $sl(2)$ states is 
simple. Recall first that for  the models (\ref{ram}) 
with $m=2$, the interpretation is:
\beq
{\rm state}(\af_1)  \longleftrightarrow 
|+\frac{1}{2}\rangle_{\af_1}\quad ,\quad
{\rm state}(\af_2)  \longleftrightarrow |-\frac{1}{2}\rangle_{\af_2}
\eeq
This yields the following identifications when two spin-$\frac{1}{2}$ 
spaces are fused:
\beq
{\rm state}(\af_2\af_1)  \longleftrightarrow |0\rangle_{\af_2\af_1}\quad
, \quad n_1\, n_2
\eeq 
for the spin 0 representation obtained from $\pi^{(1)}$, and 
\bea
& & {\rm state}(\af_1\be_1)  \longleftrightarrow |+1\rangle_{\af_1\be_1}
\quad , \quad (n_1)^2 \\
& & {\rm state}(\af_1\af_2)  \longleftrightarrow |0\rangle_{\af_1\af_2}
\quad , \quad n_1\, n_2 \\
& & {\rm state}(\af_2\be_2)  \longleftrightarrow |-1\rangle_{\af_2\be_2}
\quad , \quad (n_2)^2 
\eea
for the spin 1 representation obtained from $\pi^{(2)}$.
The numbers on the right are the number of states of the corresponding type.
Whereas the numbers of copies of the states $|\pm 1\rangle$ are
uncorrelated, $(n_1)^2$ and $(n_2)^2$, the number of states
of type $|0\rangle$ has to be $n_1\, n_2$. 
Thus  a naive trial to obtain the multistate version  
of the spin-1$\times$spin-1 model would have failed. Fusion gives the 
correct answer.  

The first derivative at $\la=0$ of  $\check{R}^{(2)}_{<12><34>}(\la)=
\CP_{13}\CP_{24} R^{(2)}_{<12><34>}(\la)$, for
the matrix  of (\ref{rm11}), gives the hamiltonian density $H_{ii+1}$
for the multistate  version of the spin-1 model.
The $n_1=n_2=1$ Hamiltonian is also  known as the Fateev-Zamolodchikov 
model \cite{fzmod}.

\section{Concluding remarks}

The multiplicity $A_{m-1}$ models  were shown to allow fusion and
multistate models were  obtained for higher dimensional representations.
The spin-0 and spin-1 models were derived explicitly. This provided in 
particular the generalization  of the Fateev-Zamolodchikov model. 

The underlying $sl(m)$ structure of these models was justified in 
\cite{amp}. It provides a natural way to 
label them in terms of $sl(m)$ representations. This language has been used 
throughout the paper and in particular in  section \ref{sl2}.
I recall here
the main points in view of an extension to the higher dimensional models
just obtained. 
The matrices  (\ref{ram}) have the  following  structure:
$R(\la)=\sum_k g_k(\la) \CO_k$, where all the spectral parameter dependence 
is contained in the functions $g_k(\la)$. The operators $\CO_k$ may depend
on the other parameters. Replacing $R$ in the YBE's it is required to 
satisfy,  and identifying the
coefficients of the linearly independent functions yield a set of equations
for the operators $\CO_k$, where the spectral parameters do not appear. 
In the case of the $\check{\CO}_k$ of (\ref{ram}) one  has simple 
$n_i$-independent trigonometric  functions,  and obtains 
the Hecke algebra, for all values of the $n_i$'s, and
in particular for $n_i=1$. This and the fact that the operators have the same
structure as for $n_i=1$ shows that one has the natural generalization.
The algebraic Bethe Ansatz was also found to be  based on the
Dynkin diagram of  $A_{m-1}$. Thus the 
functional {\it and} operatorial structures
of $R(\la)=\sum_k g_k(\la) \CO_k$ are the same as for $n_i=1$.
The quite general  fusion construction reviewed in section 
\ref{fusionsec} preserves   the functional and algebraic structures.
Therefore one again has the natural multistate generalization for 
the higher dimensional representations. 

The rational limit of all the matrices considered above exist and provide 
rational  solutions to the various Yang-Baxter equations.
This limit is obtained by rescaling $\la$ to $\gamma \la$, and 
dividing  by  an appropriate power of $\gamma$ or $\sin\gamma$  in 
order to obtain a finite limit. 

The  models (\ref{ram}) have extended symmetries \cite{amp}:
$sl(n_1)\oplus ...\oplus sl(n_m)\oplus u(1)
\oplus ... \oplus u(1)$. It is obvious that the higher dimensional 
models inherit similar symmetries.
These are related to the 
existence of many states of a same type, as seen in the 
spin-0  and spin-1 examples. Most (but not all) diagonal operators  
commute with the $R$-matrix, as they either correspond to $u(1)$ charges of
states of the same type, or to realizations of $S^z$. 
The non-diagonal operators which commute with 
the $R$-matrix are those which acts within the space spanned by 
a state and its copies. 

These symmetries will also be reflected in a diagonalization by Bethe
Ansatz, as happened for the fundamental models \cite{amp}.
The diagonalization makes use of the standard techniques associated
with higher dimensional representations. In particular, it uses 
fusion to find relations between the various transfer matrices.

\end{document}